\documentclass{elsart}
\usepackage{amsmath}
\begin{document}
\begin{frontmatter}
\title{Remarks on the method of comparison 
equations (generalized WKB method) and the generalized Ermakov-Pinney 
equation}
\author[BO,MO]{Alexander Kamenshchik},
\ead{kamenshchik@bo.infn.it}
\author[BO]{Mattia Luzzi} and
\ead{luzzi@bo.infn.it}
\author[BO]{Giovanni Venturi}
\ead{armitage@bo.infn.it}
\address[BO]{Dipartimento di Fisica and INFN, via Irnerio 46, 40126 Bologna, Italy}
\address[MO]{L.D. Landau Institute for Theoretical Physics, Kosygin str. 2, 119334 Moscow, Russia}
\begin{abstract}
The connection between the method of comparison equations 
(generalized WKB method) and the Ermakov-Pinney equation is established.
A perturbative scheme of solution of the generalized Ermakov-Pinney 
equation is developed and is applied to the construction of perturbative
series for second-order differential equations with and without turning 
points. 
\end{abstract} 
\end{frontmatter}
\section{Introduction: the general case}
\label{intro}
The WKB method is one of the main tools in quantum mechanics~\cite{Schiff,Berry}
and quantum field theory~\cite{DeWitt}.
From the mathematical point of view this method coincides with 
the construction of the asymptotic expansions for the solution 
to the second-order 
differential equations with the small coefficient of the 
second derivative term~\cite{Olver,Dinglebook}.
In this note we consider an interesting modification of the 
WKB method - the so called method of comparison Eqs.~\cite{Miller,Dingle}
and analyse its relation with the Ermakov - Pinney equation~\cite{Ermakov,Pinney}.
Let us consider the second-order differential equation
\begin{eqnarray}
\frac{d^2u(x)}{dx^2} = \frac{1}{\varepsilon}\omega^2(x)u(x)
\ ,
\label{equation}
\end{eqnarray}
where $\varepsilon$ is a small parameter. 
Let us suppose that one knows the solution of another 
second-order differential equation
\begin{eqnarray}
\frac{d^2U(\sigma)}{d\sigma^2} = \frac{1}{\varepsilon}
\Omega^2(\sigma)U(\sigma)
\ .
\label{equation1}
\end{eqnarray}
In this case one can represent an {\it exact} solution of the 
Eq.~(\ref{equation}) in the form 
\begin{eqnarray}
u(x) = \left(\frac{d\sigma}{dx}\right)^{-1/2}U(\sigma)
\ ,
\label{form}
\end{eqnarray}
where the relation between the variables $x$ and $\sigma$ is given 
by the equation
\begin{eqnarray}
\omega^2(x) = \left(\frac{d\sigma}{dx}\right)^2\Omega^2(\sigma) + 
\varepsilon \left(\frac{d\sigma}{dx}\right)^{1/2}\frac{d^2}{dx^2}
\left(\frac{d\sigma}{dx}\right)^{-1/2}
\ .
\label{compare}
\end{eqnarray}
On solving Eq.~(\ref{compare}) for $\sigma(x)$ and substituting
it into Eq.~(\ref{form}) one finds the solution to Eq.~(\ref{equation}).
Eq.~(\ref{compare}) can be solved by using some iterative scheme.
The application of such a scheme is equivalent to the application of the 
WKB method or, in other words, to the construction of the uniform asymptotic
expansion  for the solution of Eq.~(\ref{equation}). 
The method of construction of the solution to Eq.~(\ref{equation}) 
by means of the solutions of Eqs.~(\ref{equation1}) and (\ref{compare}) 
is called the method of comparison equations and the function 
$\Omega^2(\sigma)$ is called the comparison function~\cite{Dingle}. 
\par
The iterative scheme of solution of Eq.~(\ref{compare}) depends essentially 
on the form of the comparison function $\Omega(\sigma)$.  
A reasonable approach consists in the elimination of  
$\sigma$ from Eq.~(\ref{compare}) and its reduction to a form,
where the only unknown function is $\left({d\sigma}/{dx}\right)$.
The equation obtained for the case of the simple comparison function 
$\Omega^2(\sigma) = 1$ coincides with the Ermakov-Pinney equation, this  will 
be explicitly shown in the next section. For more complicated forms
of the comparison function $\Omega^2(\sigma)$ useful for the description 
of various physical problems starting with the classical problem of
tunneling  for non-relativistic quantum mechanics~\cite{Schiff,Berry,Langer}
and ending with the WKB-type analysis of the linearised equations for 
cosmological perturbations~\cite{cosmol-pert,WKB-MCE}, the corresponding equation
will acquire a more intricate form and we shall call it a generalised 
Ermakov-Pinney equation. The role of an unknown function in this equation
is played by the function
\begin{eqnarray}
y(x) \equiv    
\left(\frac{d\sigma}{dx}\right)^{-1/2}
\ ,
\label{y-define}
\end{eqnarray}
while the equation itself has the form 
\begin{eqnarray}
K\left(y,\frac{dy}{dx}\right) = \varepsilon L\left(y,\frac{dy}{dx},
\frac{d^2y}{dx^2},\frac{d^3y}{dx^3}\right)
\ ,
\label{gen-eq}
\end{eqnarray}
where the concrete forms of the functions $K$ and $L$ will be defined 
in the next sections. We shall represent the function $y(x)$ as a series
\begin{eqnarray}
y = \sum_{n=0}^{\infty} y_n \varepsilon^n
\ ,
\label{asymp}
\end{eqnarray}  
and the general recurrence relation connecting 
the different coefficient functions
$y_n$ can be represented as 
\begin{eqnarray}
\left.\left(\frac{d^nK}{d\varepsilon^n} - \frac{d^{n-1}L}{d\varepsilon^{n-1}}
\right)\right|_{\varepsilon = 0} = 0
\ .
\label{gen-rec}
\end{eqnarray}
To implement this formula one can use the combinatorial relation~\cite{Grad}
\begin{eqnarray}
\frac{d^n(uv)}{dx^n} = (u+v)^{(n)}
\ ,
\label{Grad}
\end{eqnarray}
where on the right-hand side one has a binomial expression wherein 
the powers are replaced by the derivatives.
When the function $y(x)$  is constructed up to some level of approximation
one can, in principle,  find $\sigma(x)$ and substitute for it into 
Eq.~(\ref{form}). 
\par
We shall consider the implementation of this scheme 
for four choices of the comparison function: $\Omega^2(\sigma) = 1,
\Omega^2(\sigma) = \sigma, \Omega^2(\sigma) = \exp(\sigma) -1$ and 
$\Omega^2(\sigma) = \sigma^2 - a^2$. At the end of the paper we shall 
discuss the relation between the different versions of the WKB method 
and  equations of the Ermakov-Pinney type.
\section{Differential equation without turning points}
\label{NO_TP}
In the case for which the function $\omega^2(x)$ does not have zeros (turning
points) it is convenient to choose the comparison function 
$\Omega^2(\sigma) = 1$.
In this case the function $U(\sigma)$ is simply an exponent, while
Eq.~(\ref{compare}) can be rewritten as
\begin{eqnarray}
\varepsilon\frac{d^2y(x)}{dx^2} = \omega^2(x)y(x) - \frac{1}{y^3(x)}
\ ,
\label{Pinney}
\end{eqnarray}
where the function $y(x)$ is defined by Eq.~(\ref{y-define}).
The above equation is nothing but the well known Pinney or Ermakov-Pinney
Eq.~\cite{Ermakov,Pinney}, which can be solved perturbatively with
respect to the parameter $\varepsilon$~\cite{Lewis}. 
We shall give here some general formulae for such a solution.
It is convenient to rewrite Eq.~(\ref{Pinney}) as 
\begin{eqnarray}
\omega^2 y^4 - 1 = \varepsilon y^3\ddot{y}
\ ,
\label{Pinney1}
\end{eqnarray}
where a ``dot'' denotes the derivative with respect to $x$. 
We shall search for the solution of Eq.~(\ref{Pinney1}) in the form~(\ref{asymp}).
The zero-order solution of Eq.~(\ref{Pinney1}) is 
\begin{eqnarray}
y_0 = \omega^{-1/2}
\ ,
\label{zero-order}
\end{eqnarray}
and the general recurrence relation following from Eq.~(\ref{Pinney1}) 
for $n \geq 1$ is
\begin{eqnarray}
y_n &=& -\frac{1}{4y_0^3}\sum_{k_1 = 0}^{n-1}
\sum_{k_2 = 0}^{n-1}
\sum_{k_3 = 0}^{n-1}
\sum_{k_4 = 0}^{n-1}
\delta\left[n-\left(k_1+k_2+k_3+k_4\right)\right]y_{k_1}y_{k_2}y_{k_3}y_{k_4}\nonumber
\\
&+&\frac{1}{4\omega^2 y_0^3}
\sum_{k_1 = 0}^{n-1}
\cdots
\sum_{k_4 = 0}^{n-1}
\delta\left[n-1-\left(k_1+k_2+k_3+k_4\right)\right]
\ddot{y}_{k_1}y_{k_2}y_{k_3}y_{k_4}
\ ,
\label{recurrent}
\end{eqnarray}
where $\delta()$ is the Kronecker delta symbol. 
It is now easy to obtain from the general expression~(\ref{recurrent}) 
expressions for particular values of $y_n$
\begin{subequations}
\begin{eqnarray}
y_1 &=& \frac{\ddot{y}_0}{4\omega^2}
\label{y1}
\\
y_2 &=& -\frac{6y_1^2}{4y_0} + \frac{\ddot{y}_1}{4\omega^2} 
+\frac{3\ddot{y_0}y_1}{4\omega^2 y_0}
\label{y2}
\\
y_3 &=& -\frac{y_1^3}{y_0^2} - \frac{3y_2 y_1}{y_0}
 +\frac{\ddot{y}_2}{4\omega^2} \nonumber \\
&+&\frac{3\ddot{y}_0 y_2}{4\omega^2 y_0} +
\frac{3\ddot{y}_1 y_1}{4\omega^2 y_0} +\frac{3\ddot{y}_0 y_1^2}
{4\omega^2 y_0^2}
\ .
\label{y3}
\end{eqnarray}
\end{subequations}
Let us end  by noting that the standard WKB approximation 
corresponds to the case discussed in this section and is of course only 
valid away from turning points. 
\section{Turning point: the Langer solution}
\label{langer_TP}
In the case of the presence of a linear zero in the 
function $\omega^2(x)$ one can use 
the Langer solution~\cite{Langer}. In terms of the method of comparison 
equations it means that one chooses the comparison function 
$\Omega^2(\sigma) = \sigma$. In this case Eq.~(\ref{compare}) becomes
\begin{eqnarray}
\omega^2(x) = \left(\frac{d\sigma}{dx}\right)^2\sigma + 
\varepsilon \left(\frac{d\sigma}{dx}\right)^{1/2}\frac{d^2}{dx^2}
\left(\frac{d\sigma}{dx}\right)^{-1/2}
\ .
\label{compare1}
\end{eqnarray}
Dividing this equation by $\left({d\sigma}/{dx}\right)^2$ 
and differentiating the result with respect to $x$ we get an equation
which depends only on the derivative $\dot{\sigma}$ and not on the function
$\sigma$. Such an equation can be rewritten in the form 
\begin{eqnarray}
2\dot{\omega}\omega y^6 + 4\omega^2\dot{y}y^5 = 
1 + \varepsilon\left(3\ddot{y}\dot{y}y^4
+y^{(3)}y^5\right)
\ ,
\label{Langer}
\end{eqnarray}
where the function $y(x)$, as  in the preceding section,  
is defined by Eq.~(\ref{y-define}).
Again we shall search for the solution of Eq.~(\ref{Langer}) in the form~(\ref{asymp}).
The equation for $y_0$ is
\begin{eqnarray}
2\dot{\omega}\omega y_0^6 + 4\omega^2\dot{y_0}y_0^5 = 
\left(\omega^2 y_0^4\right)^{.}y_0^2 = 1
\ .
\label{Langer0}
\end{eqnarray}
This equation my be rewritten as
\begin{eqnarray}
\sqrt{\omega^2 y_0^4}\,\frac{d}{dx}\left(\omega^2 y_0^4\right) = \omega
\ ,
\end{eqnarray}
and integrated 
\begin{eqnarray}
\omega^2 y_0^4 = \left(\frac32\int \omega  dx\right)^{2/3}
\ .
\label{langer01}
\end{eqnarray}
Finally
\begin{eqnarray}
y_0 = \frac{1}{\omega^{1/2}} \left(\frac32\int \omega dx\right)^{\frac16}
\ .
\label{Langer02}
\end{eqnarray}
Comparing the terms arising in Eq.~(\ref{Langer}) as  coefficients 
of $\varepsilon^n$ with $n \geq 1$ we have
\begin{eqnarray}
4\omega^2 y_0^5 \dot{y}_n + 
\left(12\dot{\omega}\omega y_0^5 + 20\omega^2\dot{y}_0y_0^4\right)y_n =
4\omega^{-1/2}\left(\omega^{3}y_0^5y_n\right)^{.} = F_n
\ ,
\label{Langern}
\end{eqnarray}
where $F_n$ contains the terms which depend on $y_k,0\leq  k \leq n-1$, 
and can be written as
\begin{eqnarray}
\!\!\!\!\!\!\!\!\!\!\!
F_n &\!=\!& \sum_{k_1=0}^{n-1}
\cdots
\sum_{k_6=0}^{n-1}
\left\{
\delta\left[n\!-\!\left(k_1\!+\!k_2\!+\!k_3\!+\!k_4\!+\!k_5\!+\!k_6\right)\right]
\left(4\omega^2\dot{y}_{k_1}y_{k_2}-2\dot{\omega}\omega y_{k_1}y_{k_2}\right)
\right.
\nonumber \\
\!\!\!\!\!\!\!\!\!\!\!
&\!+\!&\left.\delta[n\!-\!1\!-\!(k_1\!+\!k_2\!+\!k_3\!+\!k_4\!+\!k_5\!+\!k_6)]
\left(3\ddot{y}_{k_1}\dot{y}_{k_2}
+y^{(3)}_{k_1}y_{k_2}\right)
y_{k_3}y_{k_4}y_{k_5}y_{k_6}\right\}
\ .
\label{Fn}
\end{eqnarray}
The general expression for $y_n$ is
\begin{eqnarray}
y_n =\frac{1}{4\omega^{3}y_0^5}\int dx \omega F_n
\ .
\label{Langern1}
\end{eqnarray}
Let us also give  the explicit expressions for the first terms of the
expansion~(\ref{asymp})
\begin{subequations}
\begin{eqnarray}
y_1 &=&\frac{1}{4\omega^{3}y_0^5}\int dx \omega 
\left(3\ddot{y}_0\dot{y}_0y_0^4 + y^{(3)}_0 y_0^5\right)
\label{Langer1}
\\
y_2 &=&\frac{1}{4\omega^{3}y_0^5}\int dx \omega 
\left(-30\dot{\omega}\omega y_1^2y_0^4 - 20\omega^2\dot{y}_1y_1y_0^4 
-40\omega^2\dot{y}_0y_1^2y_0^4 \right.\nonumber \\
&+&\left.3\ddot{y}_1\dot{y}_0y_0^4 + 3\ddot{y}_0\dot{y}_1y_0^4 + 
y^{(3)}_1y_0^5 + 5y^{(3)}_0y_1y_0^4\right)
\label{Langer2}
\\
y_3 &=&\frac{1}{4\omega^{3}y_0^5}\int dx \omega 
\left(-60\dot{\omega}\omega y_2y_1y_0^4 - 40\dot{\omega}\omega y_1^3y_0^3 - 
20\omega^2\dot{y}_2y_1y_0^4
\right.\nonumber \\
&-&\left.80\omega^2\dot{y}_0y_2y_1y_0^3 
+3\ddot{y}_2\dot{y}_0y_0^4 +3\ddot{y}_0\dot{y}_2y_0^4 + 
12\ddot{y}_0\dot{y}_0y_2y_0^3 
+3\ddot{y}_1\dot{y}_1y_0^4
\right.\nonumber \\
&+&\left.12\ddot{y}_1\dot{y}_0y_1y_0^3
+12\ddot{y}_0\dot{y}_1y_1y_0^3
+18\ddot{y}_0\dot{y}_0y_1^2y_0^2
+y^{(3)}_2y_0^5
\right.\nonumber \\
&+&\left.5y^{(3)}_0y_2y_0^4
+5y^{(3)}_1y_1y_0^4
+10y^{(3)}_0y_1^2y_0^3\right)
\ .
\label{Langer3}
\end{eqnarray}
\end{subequations}
Before closing this section we may  compare  our results 
with some formulae, obtained by Dingle~\cite{Dingle}.
In the vicinity of the turning point, which we choose as $x=0$, 
the function $\omega^2(x)$ can be represented as 
\begin{eqnarray}
\omega^2(x) = \gamma_1 x + \frac{\gamma_2}{2}x^2 + \frac{\gamma_3}{3!}x^3
+ \frac{\gamma_4}{4!}x^4+ \cdots
\label{gamma-turn}
\end{eqnarray}
while the function $\sigma(x)$ looks like  
\begin{eqnarray}
\sigma(x) = \sigma_0 + \sigma_1 x + \frac{\sigma_2}{2}x^2 + \frac{\sigma_3}{3!}x^3
+ \frac{\sigma_4}{4!}x^4 + \cdots
\label{sigma-turn}
\end{eqnarray}
In reference~\cite{Dingle} the first five coefficients $\sigma_0,\sigma_1,\sigma_2,
\sigma_3$ and $\sigma_4$, as functions of the coefficients 
$\gamma_1,\gamma_2,\gamma_3$ and $\gamma_4$, were obtained in the 
lowest approximation by using a system of recurrence relations. We can 
reproduce
all these formulae by only using the  general formula~(\ref{Langer02}). 
Indeed, on writing
\begin{eqnarray}
\frac{d\sigma}{dx} = \frac{1}{y^2} = \frac{\omega(x)}
{\left[\frac32\int \omega(x)\right]^{1/3}}
\label{Langer-turn}
\end{eqnarray}
using expansions~(\ref{gamma-turn}) and (\ref{sigma-turn}) and 
comparing the coefficients of different powers of $x$ we get the formulae
\begin{eqnarray}
\sigma_0 &=& 0 \nonumber \\
\sigma_1 &=& \gamma_1^{1/3} \nonumber \\
\sigma_2 &=& \frac15\gamma_1^{-2/3}\gamma_2 \nonumber \\
\sigma_3 &=& \frac17\gamma_1^{-2/3}\gamma_3 - \frac{12}{175}\gamma_1^{-5/3}
\gamma_2^2\nonumber \\
\sigma_4 &=& \frac19\gamma_1^{-2/3}\gamma_4 -
\frac{44}{315}\gamma_1^{-5/3}\gamma_2\gamma_3 + \frac{148}{2625}
\gamma_1^{-8/3}\gamma_2^3
\ ,
\label{sigma-turn1}
\end{eqnarray}
which coincide with those obtained in~\cite{Dingle}. The value of $\sigma_0$ comes directly
from Eq.~(\ref{compare1}) ignoring the $\epsilon$-term and using the
relations~(\ref{gamma-turn}) and (\ref{sigma-turn}).
\par
We can also find the $\varepsilon$-dependent corrections to the function
$\sigma(x)$ in the neighbourhood of the point $x = 0$ by using the
results~(\ref{sigma-turn1}) and the recurrence relations~(\ref{Langer1}) etc.
We shall only give here the first correction, proportional 
to $\varepsilon$, to the coefficients $\sigma_0$ and $\sigma_1$.
The value of $\sigma_0$ comes directly from Eq.~(\ref{compare1}) using the
relations~(\ref{gamma-turn}), (\ref{sigma-turn}) and (\ref{sigma-turn1})
\begin{eqnarray}
\sigma_0 = \varepsilon\left(\frac{\gamma_3\gamma_1^{-5/3}}{14}
-\frac{9\gamma_2^2\gamma_1^{-8/3}}{140}\right)
\ .
\label{sigma0-next}
\end{eqnarray}
In order to obtain $\sigma_1$ we introduce into the definition of the 
function $y(x)$, Eq.~(\ref{y-define}), the expansion~(\ref{sigma-turn}) and
\begin{eqnarray} 
\!\!\!\!\!\!\!\!\!\!
y_0\!=\!\sigma_1^{-1/2}\left[1\!-\!\frac{\sigma_2}{2\sigma_1}x
\!+\!\left(\frac{3\sigma_2^2}{8\sigma_1^2}\!-\!\frac{\sigma_3}{4\sigma_1}\right)x^2
\!+\!\left(\frac{3\sigma_2\sigma_3}{8\sigma_1^2}\!-\!\frac{\sigma_4}{12\sigma_1}
\!-\!\frac{5\sigma_2^3}{16\sigma_1^3}\right)x^3\right]\!+\cdots
\label{y0-neig}
\end{eqnarray}
Now, on using the recurrence formula~(\ref{Langer1}) we find 
the leading (constant) term in the expression for $y_1$
\begin{eqnarray}
y_1 = \frac{1}{\gamma_1}\left(-\frac{1}{2}\sigma_2^3\sigma_1^{-7/2}
+\frac{1}{2}\sigma_3\sigma_2\sigma_1^{-5/2}-\frac{\sigma_4\sigma_1^{-3/2}}{12}
\right) + \cdots
\label{y1-neig}
\end{eqnarray}
From the formula
\begin{eqnarray}
\frac{d\sigma}{dx} = \frac{1}{y^2} = \frac{1}{(y_0 + \varepsilon y_1)^2}
\label{sigma-y}
\end{eqnarray}
it is easy to find the first correction to the first coefficient 
$\sigma_1$ in the expansion~(\ref{sigma-turn})
\begin{eqnarray}
\sigma_1 = \gamma_1^{1/3} - 2\varepsilon \frac{y_1}{y_0^3}
\ .
\label{sigma-cor}
\end{eqnarray}
Substituting into Eq.~(\ref{sigma-cor}) the leading term of 
$y_1$ from Eq.~(\ref{y1-neig}), of $y_0$ from~(\ref{y0-neig}) and 
using the explicit expressions for the coefficients $\sigma_i$ from 
Eq.~(\ref{sigma-turn1}) we finally obtain 
\begin{eqnarray}
\sigma_1 =
\gamma_1^{1/3}
+ \varepsilon\left(\frac{7}{225}\gamma_2^3
\gamma_1^{-11/3} -\frac{7}{135}\gamma_3\gamma_2\gamma_1^{-8/3} 
+\frac{1}{54}\gamma_4\gamma_1^{-5/3}\right)
\ .
\label{sigma-cor1}
\end{eqnarray}
The results~(\ref{sigma0-next}) and (\ref{sigma-cor1}) was found by
Dingle~\cite{Dingle}, but in his article there were two sign errors in the second line
of equation (69).
\section{Turning point: more complicated comparison function}
\label{cosmpert_TP}
Let us consider the differential Eq.~(\ref{equation}),
when the function $\omega^2(x)$ has a linear turning point, but
instead of Langer's comparison function $\Omega^2(\sigma) = \sigma$
consider a more complicated comparison function, also having linear turning 
point, namely
\begin{eqnarray}
\Omega^2(\sigma) = e^{\sigma}-1
\ .
\label{gammanew}
\end{eqnarray}
Now, Eq.~(\ref{compare}) will have the form
\begin{eqnarray}
\omega^2(x) = \left(\frac{d\sigma}{dx}\right)^2\left(e^{\sigma}-1\right) + 
\varepsilon \left(\frac{d\sigma}{dx}\right)^{1/2}\frac{d^2}{dx^2}
\left(\frac{d\sigma}{dx}\right)^{-1/2}
\ .
\label{compare2}
\end{eqnarray}
Isolating the function $e^{\sigma}$, taking its logarithm and 
differentiating it with respect to $x$ we arrive to the equation which
involves only the derivative $\dot{\sigma}$. This equation can be written
down as
\begin{eqnarray}
\omega^2 y^4 -2\dot{\omega}\omega y^6 -4\omega^2\dot{y}y^5 + 1 
= \varepsilon\left(\ddot{y}y^3 - 3\ddot{y}\dot{y}y^4 - y^{(3)}y^5\right)
\ .
\label{nonLanger}
\end{eqnarray}
To find the function $y_0$ one should solve the equation
\begin{eqnarray}
\omega^2 y_0^4 -2\dot{\omega}\omega y_0^6 -4\omega^2\dot{y}_0y_0^5 + 1 = 0
\ .
\label{nonLanger0}
\end{eqnarray} 
On introducing the new variable 
\begin{eqnarray}
z \equiv \omega^2 y_0^4
\label{z-define}
\end{eqnarray}
one can rewrite Eq.~(\ref{nonLanger0}) as 
\begin{eqnarray}
\dot{z} = \frac{(z+1)\omega}{z^{1/2}}
\ .
\label{z-eq}
\end{eqnarray}
It is also convenient  to introduce the variable
\begin{eqnarray}
v \equiv \sqrt{z}
\ .
\label{v-define}
\end{eqnarray}
Eq.~(\ref{z-eq}) now becomes
\begin{eqnarray}
\frac{2v^2\dot{v}}{1+v^2} = \omega
\label{v-eq}
\end{eqnarray}
which can be integrated, obtaining 
\begin{eqnarray}
2v -2\arctan v = \int \omega dx
\ .
\label{v-sol}
\end{eqnarray}
Finally, for $y_0$ we have the implicit representation
\begin{eqnarray}
y_0^2 - \frac{1}{\omega} \arctan \omega y_0^2 =
\frac{1}{2\omega}\int \omega dx
\ .
\label{nonLanger01}
\end{eqnarray}
We may now write the equation defining the recurrence relation
for $y_n, n\geq 1$
\begin{eqnarray}
-4\omega^2 y_0^5 \dot{y}_n + \left(4\omega^2 y_0^3 - 20\omega^2 \dot{y}_0y_0^4
-12\dot{\omega}\omega y_0^5\right)y_n = F_n,
\label{nonLangern}
\end{eqnarray}
where 
\begin{eqnarray}
F_n &\!=\!& -\omega^2
\sum_{k_1=0}^{n-1}
\cdots
\sum_{k_4=0}^{n-1}
\delta\left[n\!-\!\left(k_1\!+\!k_2\!+\!k_3\!+\!k_4\right)\right]y_{k_1}y_{k_2}y_{k_3}y_{k_4}\nonumber \\
&\!+\!&2\dot{\omega}\omega
\sum_{k_1=0}^{n-1}
\cdots
\sum_{k_6=0}^{n-1}
\delta\left[n\!-\!\left(k_1\!+\!k_2\!+\!k_3\!+\!k_4\!+\!k_5\!+\!k_6\right)\right]y_{k_1}\cdots y_{k_6}
\nonumber \\
&\!+\!&4\omega^2
\sum_{k_1=0}^{n-1}
\cdots
\sum_{k_6=0}^{n-1}
\delta\left[n\!-\!\left(k_1\!+\!k_2\!+\!k_3\!+\!k_4\!+\!k_5\!+\!k_6\right)\right]
\dot{y}_{k_1}y_{k_2}\cdots y_{k_6}
\nonumber \\
&\!+\!&\sum_{k_1=0}^{n-1}
\cdots
\sum_{k_4=0}^{n-1}
\delta\left[n\!-\!1\!-\!\left(k_1\!+\!k_2\!+\!k_3\!+\!k_4\right)\right]
\ddot{y}_{k_1}y_{k_2}y_{k_3}y_{k_4}\nonumber \\
&\!-\!&3\sum_{k_1=0}^{n-1}
\cdots
\sum_{k_6=0}^{n-1}
\delta\left[n\!-\!1\!-\!\left(k_1\!+\!k_2\!+\!k_3\!+\!k_4\!+\!k_5\!+\!k_6\right)\right]
\ddot{y}_{k_1}\dot{y}_{k_2}y_{k_3}\cdots y_{k_6}
\nonumber \\
&\!-\!&\sum_{k_1=0}^{n-1}
\cdots
\sum_{k_6=0}^{n-1}
\delta\left[n\!-\!1\!-\!\left(k_1\!+\!k_2\!+\!k_3\!+\!k_4\!+\!k_5\!+\!k_6\right)\right]
y^{(3)}_{k_1}y_{k_2}\cdots y_{k_6}
\ .
\label{nonLangerF}
\end{eqnarray}
The solution for $y_n$ is
\begin{eqnarray}
y_n = -\frac{1}{4\omega^{3}y_0^5G}\int dx\,\omega G F_n
\ ,
\label{nonLangernsol}
\end{eqnarray}
where
\begin{eqnarray}
G(x) = \exp\left(-\int^{x}\frac{dx'}{y_0^2(x')}\right)
\ .
\label{G-define}
\end{eqnarray}
Here we give the explicit expressions for the first two functions $F_1$ and $F_2$
\begin{subequations}
\begin{eqnarray}
\!\!\!\!\!
F_1 &=& \ddot{y}_0y_0^3 - 3\ddot{y}_0\dot{y}_0y_0^4 - y^{(3)}_0
y_0^5
\ ,
\label{nonLangerF1}\\
\!\!\!\!\!
F_2 &=& -5\omega^2 y_1^2 y_0^2 + 30\dot{\omega}\omega y_1^2y_0^4
+20\omega^2\dot{y}_1y_1y_0^4
+40\omega^2\dot{y}_0y_1^2y_0^3 
+\ddot{y}_1y_0^5
\nonumber \\
\!\!\!\!\!
&+&5\dot{y}_0y_1y_0^4
-3\ddot{y}_1\dot{y}_0y_0^4
-3\ddot{y}_0\dot{y}_1y_0^4
-12\ddot{y}_0\dot{y}_0y_1y_0^3
-y^{(3)}_1y_0^5
-5y^{(3)}_0y_1y_0^4
\ .
\label{nonLangerF2}
\end{eqnarray}
\end{subequations}
\section{Two turning points}
\label{2_TP}
Let us now consider the differential Eq.~(\ref{equation}), when the 
function $\omega^2(x)$ has two turning points. This situation, describing 
the tunneling through a potential barrier was considered in~\cite{Miller,Berry}.
In this case the comparison function can be chosen as
\begin{eqnarray}
\Omega^2(\sigma) = \sigma^2 - a^2
\ .
\label{comp-two}
\end{eqnarray}
The solutions  $U(\sigma)$ of Eq.~(\ref{equation1}) are the well-known 
parabolic functions~\cite{Berry}. Substituting  $\Omega^2(\sigma)$
of Eq.~(\ref{comp-two}) into Eq.~(\ref{compare}) one can isolate 
$\sigma$ and after the subsequent differentiation one can write down the
equation for the function $y(x)$
\begin{eqnarray}
2\sqrt{\omega^2 y^4 + a^2 + \varepsilon \ddot{y}y^3} = 
2\dot{\omega}\omega y^6 +4\omega^2\dot{y}y^5 + 
\varepsilon\left(3\ddot{y}\dot{y}y^4
+y^{(3)}y^5\right)
\ .
\label{comp-two1}
\end{eqnarray}
The lowest-approximation $y_0$ can be found from the equation
\begin{eqnarray}
2\sqrt{\omega^2 y^4 + a^2} = 
2\dot{\omega}\omega y^6 +4\omega^2\dot{y}y^5
\ .
\label{comp-two2}
\end{eqnarray}
This equation can be integrated and the solution $y_0$ can be written down 
in the implicit form
\begin{eqnarray}
\omega y_0^3 -\frac{a^2}{\omega}{\rm arcsinh}
\frac{\omega y_0}{a} = \frac{2}{\omega}\int\omega dx
\ .
\label{implicit}
\end{eqnarray}
To get the recurrence relations for the functions $y_n, n \geq 1$ it is 
convenient to take the square of Eq.~(\ref{comp-two1})
\begin{eqnarray}
\!\!\!\!\!\!\!
&&4\dot{\omega}^2\omega^2 y^{12} + 16\omega^4\dot{y}^2y^{10} 
+ 16\dot{\omega}\omega^3\dot{y}y^{11}-4\omega^2 y^4 -4a^2
=
\nonumber \\
\!\!\!\!\!\!\!
&&\quad\quad\varepsilon(4\ddot{y}y^3-12\dot{\omega}\omega\ddot{y}\dot{y}y^{10}
-24\omega^2 \ddot{y}\dot{y}^2y^9 - 4\dot{\omega}\omega
y^{(3)}y^{11}
-8\omega^2y^{(3)}\dot{y}y^{10})
\nonumber \\
\!\!\!\!\!\!\!\
&&\quad\quad-\varepsilon^2(9\ddot{y}^2\dot{y}^2y^8+{y^{(3)}}^2y^{10}
+6y^{(3)}\ddot{y}\dot{y}y^9)
\ .
\label{comp-two3}
\end{eqnarray}
On comparing the terms containing the $n^{\rm th}$ powers of the small parameter 
$\varepsilon$ one obtains
\begin{eqnarray}
f \dot{y}_n + g y_n = F_n
\ ,
\label{rec-two}
\end{eqnarray}
where
\begin{subequations}
\begin{eqnarray}
f &=& 32\omega^4\dot{y}_0y_0^{10} + 16\dot{\omega}\omega^3 y_0^{11}
\\
\label{f-def}
g &=& 48\dot{\omega}^2\omega^2 y_0^{11} + 160\omega^4\dot{y}_0^2y_0^9
+176\dot{\omega}\omega^3\dot{y_0}y_0^{10}-16\omega^2 y_0^3
\\
\label{g-def}
F_n &=& \sum_{k_1=0}^{n-1}\cdots\sum_{k_{12}=0}^{n-1}
\delta\left[n-\left(k_1+\cdots+k_{12}\right)\right]\nonumber\\
&\times& 4\left(\dot{\omega}^2\omega^2 y_{k_1}y_{k_2}
+16\omega^4\dot{y}_{k_1}\dot{y}_{k_2}
+16\dot{\omega}\omega^3\dot{y}_{k_1}y_{k_2}\right)y_{k_3}\cdots y_{k_{12}}\nonumber \\
&-&4\omega^2\sum_{k_1=0}^{n-1}\cdots\sum_{k_4=0}^{n-1}
\delta\left[n-\left(k_1+\cdots+k_4\right)\right]y_{k_1}\cdots y_{k_4}\nonumber\\
&-&\varepsilon\left\{-4 
\sum_{k_1=0}^{n-1}\cdots\sum_{k_4=0}^{n-1}
\delta\left[n-1-\left(k_1+\cdots+k_4\right)\right]\ddot{y}_{k_1}y_{k_2}y_{k_3} y_{k_4}\right.\nonumber\\
&+&\left.
\sum_{k_1=0}^{n-1}\cdots\sum_{k_{12}=0}^{n-1}
\delta\left[n-1-\left(k_1+\cdots+k_{12}\right)\right]\right.\nonumber\\
&&\quad\quad\quad\times\left.\left(12\dot{\omega}\omega\ddot{y}_{k_1}\dot{y}_{k_2}y_{k_3}+
24\omega^2\ddot{y}_{k_1}\dot{y}_{k_2}\dot{y}_{k_3}\right.\right.\nonumber\\
&&\quad\quad\quad\quad\left.\left.+4\dot{\omega}\omega y^{(3)}y_{k_2}y_{k_3}+
8\omega^2y^{(3)}_{k_1}\dot{y}_{k_2}y_{k_3}\right)y_{k_4}\cdots y_{k_{12}}\right\}\nonumber \\
&-&\varepsilon^2
\sum_{k_1=0}^{n-2}\cdots\sum_{k_{12}=0}^{n-2}
\delta\left[n-2-\left(k_1+\cdots+k_{12}\right)\right]\nonumber\\
&\times&\left(9\ddot{y}_{k_1}\ddot{y}_{k_2}\dot{y}_{k_3}\dot{y}_{k_4}
+y^{(3)}_{k_1}y^{(3)}_{k_2}
y_{k_3}y_{k_4} + 6y^{(3)}_{k_1}\dot{y}_{k_2}y_{k_3}
y_{k_4}\right) y_{k_5}\cdots y_{k_{12}}
\ .
\label{F-two}
\end{eqnarray}
\end{subequations}
The integration of Eq.~(\ref{rec-two}) gives 
\begin{eqnarray}
y_{n} = \frac{1}{G}\int dx \frac{G F_n}{f}
\ ,
\label{yntwo}
\end{eqnarray}
where 
\begin{eqnarray}
G = \exp\left(\int dx' \frac{g}{f}\right)
\ .
\label{G-two}
\end{eqnarray}
The explicit expressions for the coefficients $F_n$ are rather 
cumbersome and we shall only write down  the coefficient $F_1$
\begin{eqnarray}
F_1 &=& 48\dot{\omega}^2\omega^2 y_{1}y_0^{11}+32\dot{y}_1\dot{y}_0y_0^{10}
+160\omega^4\dot{y}_0^2y_1y_0^9\nonumber \\
&+&16\dot{\omega}\omega^3\dot{y}_1y_0^{11} 
+ 176\dot{\omega}\omega^3 \dot{y}_0y_1y_0^{10} 
-16\omega^2 y_1y_0^3 
+4\ddot{y}_0y_0^3 \nonumber \\
&-&12\dot{\omega}\omega\ddot{y}_0\dot{y}_0y_0^{10}
-24\ddot{y}_0\dot{y}_0^2y_0^9
-4\dot{\omega}\omega y^{(3)}_0y_0^{11} -
8\omega^2y^{(3)}_0\dot{y}_0y_0^{10}
\ .
\label{F1-two}
\end{eqnarray}
\section{Discussion: WKB method and the generalized Ermakov-Pinney equation}
\label{WKB_ErmPinn}
In this paper we have studied the relation between the WKB method for the 
solution of  second-order differential equations and the Ermakov-Pinney equation.
For some non-trivial applications of the WKB 
method it is found that instead of the standard form of the Ermakov-
Pinney equation one is lead to its generalizations which are rather 
different to the known generalizations of Ermakov systems. 
\par
Let us note that although the Ermakov equation was already reproduced 
in the WKB context (first of all in the important paper by Milne~\cite{Milne}),
its generalizations arising in the process of application 
of the comparison function method~\cite{Miller,Dingle} were not yet studied,
at least to the best of our knowledge. 
In this concluding section of our paper 
we shall try to give 
a short review of the history of the Ermakov-Pinney equation 
and its applications to quantum mechanics and to similar WKB-type problems
and discuss the physical significance of its non-trivial generalization
(for a more detailed account of the history of the Eramkov-Pinney equation see
e.g.~\cite{Espinosa}).
\par
The relation between the second-order linear differential equation 
\begin{eqnarray}
\frac{d^2 u}{dx^2} + \omega^2(x)u(x) = 0
\label{Ermakov}
\end{eqnarray}
and the non-linear differential equation 
\begin{eqnarray}
\frac{d^2 \rho}{dx^2} + \omega^2(x)\rho(x) = \frac{\alpha}{\rho^3(x)}
\ ,
\label{Ermakov1}
\end{eqnarray} 
where $\alpha$ is  some constant
was noticed by Ermakov~\cite{Ermakov}, who  showed 
that any two  solutions $\rho$ and $u$  of the above equations are 
connected by the formula
\begin{eqnarray}
C_1 \int \frac{dx}{u^2} + C_2 = \sqrt{C_1\frac{\rho^2}{u^2} - \alpha}
\ .
\label{Ermakov2}
\end{eqnarray}
The couple of Eqs.~(\ref{Ermakov}) and (\ref{Ermakov1}) constitute
the so called Ermakov system. An important corollary was derive~\cite{Ermakov}
from the formula~(\ref{Ermakov2}).
Namely, on having a particular solution $\rho(x)$ of Eq.~(\ref{Ermakov1}) 
one can construct the general solution of Eq.~(\ref{Ermakov})  
which is given by
\begin{eqnarray}
u(x) = c_1 \rho(x) \exp\left[\sqrt{-\alpha}\int \frac{dx}{\rho^2(x)}\right] +  
c_2 \rho(x) \exp\left[-\sqrt{-\alpha}\int \frac{dx}{\rho^2(x)}\right]
\ .
\label{Ermakov3}
\end{eqnarray}
It is easy to see that the function $\rho(x)$ plays of the role of an 
``amplitude'' of the function $u(x)$, while the integral 
$\int \left[{dx}/{\rho^2(x)}\right]$ represents some kind of ``phase''. 
Thus, it is not surprising that the Ermakov equation was re-discovered by 
Milne~\cite{Milne} in a quantum-mechanical context. On introducing the 
amplitude, obeying Eq.~(\ref{Ermakov1}) and the phase, Milne  constructed 
a formalism for the solution of the Schr\"odinger equation which was 
equivalent to the WKB method. Milne's version of the WKB technique was 
extensively used for the solution of quantum mechanical problems 
(see e.g.~\cite{Fano}). In the paper by Pinney~\cite{Pinney} the general form 
of the  solution of the Ermakov equation was presented,
while the most general expression for this solution was written down by 
Lewis~\cite{Lewis}. The equation, sometimes called Ermakov-Milne-Pinney 
equation, has also found application  in cosmology~\cite{cosmology}.   
\par
A very simple physical example rendering trasparent the derivation 
of the Ermakov-Pinney equation and its solutions was given in the paper 
by Eliezer and Gray~\cite{Gray}. The motion of a two-dimensional 
oscillator with  time-dependent frequency was considered. In this case 
the second-order linear differential Eq.~(\ref{Ermakov})
describes the projection of the 
 motion of 
this two-dimensional oscillator on one of its Cartesian coordinates,
while the Ermakov-Pinney Eq.~(\ref{Ermakov1}) describes the evolution 
of the radial coordinate $\rho$. The parameter $\alpha$ is nothing more than
the squared conserved angular momentum of the two-dimensional oscillator.
Thus, the notion of the amplitude and phase acquire in this example a simple
geoemtrical and physical meaning.  
\par
An additional generalization of the notion of Ermakov system of equations was 
suggested in the paper by Ray and Reid~\cite{Ray}. Thy consider 
the system of two 
equations  
\begin{eqnarray}
\frac{d^2 u}{dx^2} + \omega^2(x)u(x) = \frac{1}{\rho x^2}g
\left(\frac{\rho}{x}\right)
\label{Ray}
\end{eqnarray}
and  
\begin{eqnarray}
\frac{d^2 \rho}{dx^2} + \omega^2(x)\rho(x) = \frac{1}{x\rho^2(x)}
f\left(\frac{x}{\rho}\right)
\ ,
\label{Ray1}
\end{eqnarray} 
where $g$ and $f$ are arbitrary functions of their arguments.
The standard Ermakov system~(\ref{Ermakov}), (\ref{Ermakov1})
corresponds to the choice of functions $g(\xi) = 0$ and 
$f(\xi) = \alpha\xi$. One can show that the generalized Ermakov system~(\ref{Ray}),
(\ref{Ray1}) has an invariant (zero total time derivative) 
\begin{eqnarray}
I_{f,g} = \frac12\left[\phi\left(\frac{u}{\rho}\right) 
+\theta\left(\frac{\rho}{u}\right)
+\left(u\frac{d\rho}{dx}-\rho\frac{du}{dx}\right)^2\right]
\ ,
\label{invar}
\end{eqnarray}
where 
\begin{subequations}
\begin{eqnarray}
\phi\left(\frac{u}{\rho}\right) &=& 2\int^{\frac{u}{\rho}}f(x)dx
\label{invar1}
\\
\theta\left(\frac{\rho}{u}\right) &=& 2\int^{\frac{\rho}{u}}g(x)dx
\ .
\label{invar2}
\end{eqnarray}
\end{subequations}
The invariant~(\ref{invar}) establishes the connection between 
the solutions of Eqs.~(\ref{Ray}) and (\ref{Ray1}) and sometimes 
allows one to  find the solution  of one of these equations provided 
the solution of the other one is known (just as in the case of the standard
Ermakov system). However, for the case of an arbitrary couple of functions 
$f$ and $g$ a simple physical or geometrical interpretation of the 
Ermakov system analogous to that given in~\cite{Gray} is not known.
\par
In the present paper we, on one hand, have established the connection between 
the generalized WKB method (method of  comparison Eqs.~\cite{Miller,Dingle})
and the Ermakov - Pinney equation and on the other hand we have obtained a
generalization of the Ermakov - Pinney equation, different to those 
studied in the literature. 
Let us  briefly summarize  our approach.  
One has two second-order differential Eqs.~(\ref{equation}) and
(\ref{equation1}) and the solution of one of them~(\ref{equation1}) is 
well known and described. Using the previous convenient analogy~\cite{Gray},
one can say that in this case  
one has two time-dependent oscillators, evolving with two different 
time-parameters. One can try to find the solution of Eq.~(\ref{equation})
 by representing it as a known  solution of Eq.~(\ref{equation1})
multiplied by a correction factor. 
This correction factor plays the role of the prefactor
in the standard WKB approach while the known solution of 
Eq.~(\ref{equation1}) represents some kind of  generalized phase term. 
Further, the prefactor is expressed in terms of the derivative 
between two variables $x$ and $\sigma$ (or in terms of the oscillator 
analogy, between two times). 
On writing down the equation definining this factor, which we have denoted 
 by $y(x)$ (see Eq.~(\ref{y-define})) we arrive to the Eq.~(\ref{compare}) 
which could be called generalized Ermakov-Pinney equation. 
For the case when the function $\omega^2(x)$ does not have turning points
the comparison function $\Omega^2(\sigma)$ can be chosen constant
(the second oscillator has a time-independent frequency) and the 
the equation defining the prefactor becomes the standard 
Ermakov-Pinney Eq.~(\ref{Pinney}). In terms of the two-dimensional
oscillator analogy~\cite{Gray}
this means that we exclude from the equation for the radial
coordinate the dependence on the angle coordinate $\sigma$ by using its
cyclicity, i.e. the conservation of the angular moment. 
\par
In cases for which the function $\omega^2(x)$ has 
turning points, as in Secs.~\ref{langer_TP}, \ref{cosmpert_TP} and \ref{2_TP},
the comparison functions are non-constant and instead of the standard 
Ermakov-Pinney equation we have its non-trivial generalizations. The common 
feature of these generalizations consists in the fact that the corresponding
equations for the variable $y$ depend explicitly also on the parameter 
$\sigma$ which is not excluded automatically. To get a differential equation
for $y$, one should isolate the parameter $\sigma$ and subsequently 
differentiate with respect to $x$. As a result one gets a differential 
equation of higher-order for the function $y(x)$. Remarkably, for the 
perturbative solution of
these equations one can again construct  a reasonable iterative procedure.
\par
Again, it is interesting to look at the generalized Ermakov-Pinney equation
as an equation describing a two-dimensional physical system in the spirit 
of the reference~\cite{Gray}. In our case one has
\begin{eqnarray}
\frac{d^2y(x)}{dx^2} + \omega^2(x) y(x) = \frac{\Omega^2(\sigma)}{y^3(x)}
\ ,
\label{Gray}
\end{eqnarray}
where $y$ plays the role of a radial coordinate, $\sigma$ 
resembles a phase or 
an angle (e.g. the position of a hand of a clock)
and $x$ is a time parameter. It is important to notice
that according to the definition of the variable $y$~(\ref{y-define}) 
there is a relation between the radial and angle coordinates 
\begin{eqnarray}
\frac{d\sigma}{dx} = \frac{1}{y^2}
\ ,
\label{angular}
\end{eqnarray}
which 
on interpreting $\sigma$ as a phase corresponds to constant (unit) angular 
momentum.
In our case however
the right-hand side of the radial Eq.~(\ref{Gray}) 
contains an explicit dependence on the ``angle'' $\sigma$. Thus, 
the couple of Eqs.~(\ref{Gray}) and (\ref{angular})
cannot be represented as a system of equations of motion corresponding to 
some Lagrangian or Hamiltonian as in Ref.~\cite{Gray}. 
Perhaps, this system could be described 
in terms of non-Hamiltonian dissipative dynamics, but this question 
requires a further study.

\end{document}